\global\def\draftcontrol{0}

   \def\versionno{ n4rn2}

\catcode`\@=11

\expandafter\ifx\csname draftcontrol\endcsname\relax\global\def\draftcontrol{0}
\fi

{\count255=\time\divide\count255 by 60
\xdef\hourmin{\number\count255}
\multiply\count255 by-60\advance\count255 by\time
\xdef\hourmin{\hourmin:\ifnum\count255<10 0\fi\the\count255}}
\def\draftdate{\number\month/\number\day/\number\year\ \ \ \hourmin }

\newcommand\makepapertitle{\par
  \begingroup
    \renewcommand\thefootnote{\@fnsymbol\c@footnote}%
    \def\@makefnmark{\rlap{\@textsuperscript{\normalfont\@thefnmark}}}%
    \long\def\@makefntext##1{\parindent 1em\noindent
            \hb@xt@1.8em{%
                \hss\@textsuperscript{\normalfont\@thefnmark}}##1}%
     \newpage
     \global\@topnum\z@   
     \@makepapertitle
     \thispagestyle{empty}\@thanks
  \endgroup
  \setcounter{footnote}{0}%
  \global\let\thanks\relax
  \global\let\makepapertitle\relax
  \global\let\@makepapertitle\relax
  \global\let\@thanks\@empty
  \global\let\@author\@empty
  \global\let\@date\@empty
  \global\let\@title\@empty
  \global\let\title\relax
  \global\let\author\relax
  \global\let\date\relax
  \global\let\and\relax
  \def\version{\let\version\@version\@gobble}
}
\def\@makepapertitle{%
  \newpage
   \ifnum\draftcontrol=1 {}
   \version\versionno
   \vskip 3em%
   \else
   \hfill\hbox to 3cm {\parbox{4cm}{\@pubnum}\hss}%
   \vskip 3em%
   \fi
   \begin{center}%
   \let \footnote \thanks
     {\LARGE {\@title}}%
     \vskip 1.5em%
     {\normalsize
       \lineskip .5em%
       \begin{tabular}[t]{c}%
         \@author
       \end{tabular}\par}%
     \vskip 1.5em%
     {\@bstract}%
     \end{center}%
     \vskip 1.5em
     \@date%
   \par
}

\gdef\@pubnum{}
\def\pubnum#1{%
  \gdef\@pubnum{#1}}

\gdef\@bstract{}
\def\Abstract#1{%
  \gdef\@bstract{%
   \parbox{\textwidth-0pc}{%
   \centerline{\bf Abstract}\penalty1000%
\kern.2cm%
\noindent
\renewcommand\baselinestretch{1.0}%
{#1}}}
}

\def\ps@paper{\let\@mkboth\@gobbletwo%
     \ifnum\draftcontrol=1
    \def\@oddfoot{\hbox to \textwidth{\tiny \versionno \hfil\tiny\draftdate}%
    \hskip -\textwidth \hbox to \textwidth{\hfil\rm\thepage\hfil}}%
     \else\def\@oddfoot{\hbox to \textwidth{\hfil\rm\thepage\hfil}}
     \fi
     \let\@evenfoot\@oddfoot
}

\def\body{\clearpage
          \pagestyle{paper}
    }

\def\@version#1{\ifnum\draftcontrol=1
\typeout{}\typeout{#1}\typeout{}
\vskip3mm\centerline{\hbox{\fbox{\normalsize{\tt DRAFT -- #1 -- }
                   {\draftdate}}}}\vskip3mm
\fi}
\let\version\@version
\long\def\eqlabel#1{\ifnum\draftcontrol=1
                    \tag@false  
                    \tag*{(\theequation) \hbox to -0.2cm{\hspace{0cm}\small{#1}\hss}}
                    \refstepcounter{equation}
                    \edef\@currentlabel{\theequation}
                    \ltx@label{#1}          
                    \else
                    \label{#1}
                    \fi
                    }
\let\st@bibitem\@bibitem
\let\st@lbibitem\@lbibitem
\ifnum\draftcontrol=1
  \def\@bibitem#1{%
    \st@bibitem{#1}\a@@label{#1}\ignorespaces}
  \def\@lbibitem[#1]#2{%
    \st@lbibitem[#1]{#2}\a@@label{#2}\ignorespaces}
  \def\a@@label#1{%
    \gdef\a@lab{\smash{\normalfont\small#1}}
    \ifvmode
      \if@inlabel
        \global\setbox\@labels\hbox{%
          \llap{\a@lab\let\a@lab\relax
                \kern\@totalleftmargin\kern\marginparsep}%
          \box\@labels}%
      \fi
    \fi}
\fi

\documentclass[12pt,letterpaper]{article}

\usepackage{amsmath,amssymb,array,calc,epsfig,rotating,bm}
\usepackage[sort]{cite}
\usepackage{graphicx}
\usepackage{psfrag,verbatim}
\usepackage{xcolor}
\usepackage{hyperref}

\ifnum\draftcontrol=1
\fi

\renewcommand\baselinestretch{1.25}
\renewcommand\section{\@startsection {section}{1}{\z@}%
                                   {-3.5ex \@plus -1ex \@minus -.2ex}%
                                   {2.3ex \@plus.2ex}%
                                   {\normalfont\large\bfseries}}
\renewcommand\subsection{\@startsection{subsection}{2}{\z@}%
                                   {-3.25ex\@plus -1ex \@minus -.2ex}%
                                   {1.5ex \@plus .2ex}%
                                   {\normalfont\normalsize\bfseries}}
\renewcommand\subsubsection{\@startsection{subsubsection}{3}{\z@}%
                                   {-3.25ex\@plus -1ex \@minus -.2ex}%
                                   {1.5ex \@plus .2ex}%
                                   {\normalfont\normalsize\it}}
\renewcommand\paragraph{\@startsection{paragraph}{4}{\z@}%
                                   {-3.25ex\@plus -1ex \@minus -.2ex}%
                                   {1.5ex \@plus .2ex}%
                                   {\normalfont\normalsize\bf}}

\numberwithin{equation}{section}

\def\revise#1       {\raisebox{-0em}{\rule{3pt}{1em}}%
                     \marginpar{\raisebox{.5em}{\vrule width3pt\
                     \vrule width0pt height 0pt depth0.5em
                     \hbox to 0cm{\hspace{0cm}{%
                     \parbox[t]{4em}{\raggedright\footnotesize{#1}}}\hss}}}}

\newcommand{\ie}{{\it i.e.,}\ }

\def\cale         {{\cal E}}

\def\calm         {{\cal M}}
\def\caln         {{\cal N}}
\def\calo         {{\cal O}}

\def\del          {\partial}

\def\sqr#1#2{{\vcenter{\vbox{\hrule height.#2pt
 \hbox{\vrule width.#2pt height#1pt \kern#1pt
 \vrule width.#2pt}\hrule height.#2pt}}}}

\def\aa1{\phi}
\def\cc1{\psi}

\catcode`\@=12

\begin{document}


\title{\bf The ordered phase of charged $\caln=4$ SYM plasma from STU}

\date{January 11, 2025}

\author{
Alex Buchel\\[0.4cm]
\it Department of Physics and Astronomy\\ 
\it University of Western Ontario\\
\it London, Ontario N6A 5B7, Canada\\
\it Perimeter Institute for Theoretical Physics\\
\it Waterloo, Ontario N2J 2W9, Canada\\
}

\Abstract{In \cite{Buchel:2025cve} we numerically identified the ordered phase
of the charged ${\cal N}=4$ supersymmetric Yang-Mills plasma. We explain
here how this phase can be obtained analytically within the STU model
of Behrnd, Cveti\v{c}, and Sabra \cite{Behrndt:1998jd}.
}

\makepapertitle

\body

\version\versionno


Consider holographic dual of $\caln=4$ supersymmetric Yang-Mills theory \cite{Maldacena:1997re}.
The theory has $SU(4)$ R-symmetry, and it is possible to study its strongly coupled plasma
charged under the maximal Abelian subgroup $U(1)^3\subset SU(4)$ of the R-symmetry. When the chemical potentials for
all the $U(1)$ factors are the same, one possible phase of the theory is realized gravitationally as  
a Reissner-Nordstrom (RN) black hole in asymptotically $AdS_5$ space-time.
The Gibbs free energy density of this phase is given by
\begin{equation}
\Omega_{RN} =-\frac{c}{2\pi^2}\left(\alpha^4+\frac 12\alpha^2 \mu^2\right)\,,\qquad
\frac{T}{\mu}=\frac{4\alpha^2-\mu^2}{4\pi\alpha\mu}\,,
\eqlabel{omsysi}
\end{equation}
where $T$ and $\mu$ are the temperature and the chemical potential correspondingly;
$c=\frac{N_c^2}{4}$ is the central charge of the SYM, and $\alpha$ is an arbitrary auxiliary
scale\footnote{This scale can be eliminated in favor  of $\frac T\mu$.}.

In \cite{Buchel:2025cve}, a novel phase of this plasma, again with the same chemical potential
for all the $U(1)$ R-symmetry factors, was identified numerically. The phase of  \cite{Buchel:2025cve}
is an example of a {\it conformal ordered phase}\footnote{Charge neutral
conformal order was recently studied in \cite{Chai:2020zgq,Buchel:2020xdk,Buchel:2020jfs,Buchel:2020thm,Chai:2021tpt,Chaudhuri:2021dsq,Buchel:2022zxl,Chai:2021djc}.}: it extends to arbitrary high temperatures, and is characterized by the
thermal expectation value of a dimension-2 operator, with $\calo_2\propto T^2$.
In the limit
$\frac\mu T\to 0$, this ordered phase has a vanishing energy density
$\frac{{\cal E}}{T^4}\propto \frac{\mu^2}{T^2}$ and is a low entropy
density state $\frac{{s}}{T^3}\propto \frac{\mu^2}{T^2}$.
In this note we demonstrate that the ordered phase of  \cite{Buchel:2025cve}
can be understood within the class of analytic solutions of the STU model \cite{Behrndt:1998jd}.
We show that this phase actually exists for arbitrary temperatures, and at critical
temperature\footnote{This temperature signals the onset of a hydrodynamic instability in the
plasma as shown in \cite{Gladden:2024ssb}.}
$T_{crit}$,
\begin{equation}
T_{crit}=\frac{\mu}{2\pi \sqrt{2}}\,,
\eqlabel{tcrit}
\end{equation}
has the same Gibbs free energy $\Omega_{ordered}$, as that of \eqref{omsysi}.
Additionally, at fixed chemical potential $\mu$,
\begin{equation}
\frac{\Omega_{ordered}}{\mu^4}\ \lessgtr\ \frac{\Omega_{RN}}{\mu^4}\ \qquad {\rm for}\qquad  \frac T\mu\ \lessgtr\ \frac{T_{crit}}{\mu}
\eqlabel{omcomp}
\end{equation}
correspondingly. Thus, the ordered phase is a preferred one in the grand canonical ensemble at low temperatures.
In the limit $\frac{T}{\mu}\to 0$ the ordered phase has a vanishing entropy density, see \eqref{sf}.

The starting point is the STU consistent truncation of type IIB supergravity \cite{Behrndt:1998jd}: 
\begin{equation}
\begin{split}
S_{eff}=\frac{1}{2\kappa_5^2} \int_{\calm_5} \biggl(
&R-\frac 14 G_{ab}F_{\rho\sigma}^a F_{\mu\nu}^b g^{\rho\mu}g^{\sigma\nu}+
\frac{c_{abc}}{48 \sqrt{2}}\epsilon^{\mu\nu\rho\sigma\lambda}F^a_{\mu\nu}F_{\rho\sigma}^b A^c_\lambda
\\&
\qquad -G_{ab} g^{\mu\nu}\del_\mu X^a\del_\nu X^b+\sum_{a=1}^3 \frac{4}{X^a}
\biggr) \star 1\,,
\end{split}
\eqlabel{seff}
\end{equation}
where $c_{abc}$ are symmetric constants, nonzero only for distinct indices with $c_{123}=1$, $g_{\mu\nu}$ is the metric on $\calm_5$, $F^a_{\mu\nu}$ are the field strengths for the gauge fields $A^a_\mu$, $a=1\cdots 3$,
dual to conserved currents of the maximal Abelian subgroup of the $SU(4)$ $R$-symmetry of $\caln=4$ SYM. The three
real positive neutral scalar fields $X^a$  describe the deformation of $S^5$ in the uplift of $S_{eff}$ to type IIB supergravity; they are
constrained, at the level of the effective action \eqref{seff}, by
\begin{equation}
X^1 X^2 X^3=1\,.
\eqlabel{constr}
\end{equation}
The field space metric $G_{ab}$ is
\begin{equation}
G_{ab}=\frac 12 {\rm diag} \biggl((X^1)^{-2}\,,\, (X^2)^{-2}\,,\, (X^3)^{-2}\biggr)\,.
\eqlabel{const}
\end{equation}
The gravitational constant $\kappa_5$ is related to the central charge $c$ of the boundary gauge theory
as
\begin{equation}
\kappa_5^2=\frac{\pi^2}{c}=\frac{4\pi^2 }{N_c^2}\,.
\eqlabel{k5}
\end{equation}
Effective action \eqref{seff} allows for analytic solutions of black holes in asymptotically $AdS_5$ with
distinct $U(1)^3$ charges \cite{Behrndt:1998jd}, realizing the gravitational dual to charged $\caln=4$ SYM plasma:
\begin{itemize}
\item the black hole metric is
\begin{equation}
ds_5^2=-H^{-2/3} \frac{(\pi T_0)^2}{u}\ f\ dt^2 +H^{1/3}  \frac{(\pi T_0)^2}{u} d\bm{x}^2+H^{1/3} \frac{1}{4f u^2} du^2 \,,
\eqlabel{metric}
\end{equation}
where $T_0$ is an arbitrary auxiliary scale (akin to $\alpha$ in \eqref{omsysi}), the radial coordinate $u\in (0,1)$
(with $u=1$ being the black hole horizon), and the warp factors $H, f$ are ( $a=1\cdots 3$)
\begin{equation}
H\equiv \prod_a H_a\,,\qquad H_a=1+\kappa_a u\,,\qquad f=H-u^2 H(1)\,,
\eqlabel{warps}
\end{equation}
for constants $\kappa_a$ (related to $U(1)$ chemical potentials);
\item the $U(1)$ gauge fields are
\begin{equation}
A_\mu^a=\delta^t_\mu\left(\frac{1}{1+\kappa_a}-\frac{u}{H_a}\right)\pi T_0\sqrt{2\kappa_a} \prod_b(1+\kappa_b)^{1/2}\,;
\eqlabel{gauge}
\end{equation}
\item the bulk scalars are
\begin{equation}
X^a=\frac{H^{1/3}}{H_a}\,.
\eqlabel{scalars}
\end{equation}
\end{itemize}
Standard black hole thermodynamics identifies the energy density $\cale$, the entropy density
$s$, the temperature $T$, the chemical potentials $\mu_a$ and the corresponding charge densities $\rho_a$
as \cite{Harmark:1999xt}:
\begin{equation}
\begin{split}
&\cale=\frac{3}{2(2\pi N_c)^{2/3}}\ s^{4/3}\ \prod_a\left(1+\frac{8\pi^2\rho_a^2}{s^2}\right)^{1/3}\,,\qquad
s=\frac{\pi^2}{2} N_c^2 T_0^3\ \prod_a (1+\kappa_a)^{1/2}\,,\\
&\mu_a=\pi T_0\ \frac{\sqrt{2\kappa_a}}{1+\kappa_a}\ \prod_b(1+\kappa_b)^{1/2}\,,\qquad \rho_a=\frac{\pi}{8}
N_c^2 T_0^3\ \sqrt{2\kappa_a}\ \prod_b(1+\kappa_b)^{1/2}\,,\\
&T=\frac{2+\kappa_1+\kappa_2+\kappa_3-\prod_a\kappa_a}{2\prod_b(1+\kappa_b)^{1/2}}\ T_0\,.
\end{split}
\eqlabel{thermo}
\end{equation}
From the Gibbs free energy density
\begin{equation}\Omega=\cale-s T-\sum_a \mu_a\cdot \rho_a\,,
\eqlabel{g}
\end{equation}
we can readily verify the first law of thermodynamics
\begin{equation}
d\Omega=-s\cdot  dT-\sum_a\rho_a \cdot d\mu_a\,.
\eqlabel{fl}
\end{equation}

\begin{figure}[ht]
\begin{center}
\psfrag{z}[tt][][1.0][0]{{$T/\mu$}}
\psfrag{t}[tt][][1.0][0]{{$\ln( T/\mu-T/\mu|_{crit})$}}
\psfrag{x}[bb][][1.0][0]{{$\hat\Omega/\mu^4$}}
\psfrag{y}[tt][][1.0][0]{{$\ln (\Delta\hat\Omega/\mu^4)$}}
\includegraphics[width=5in]{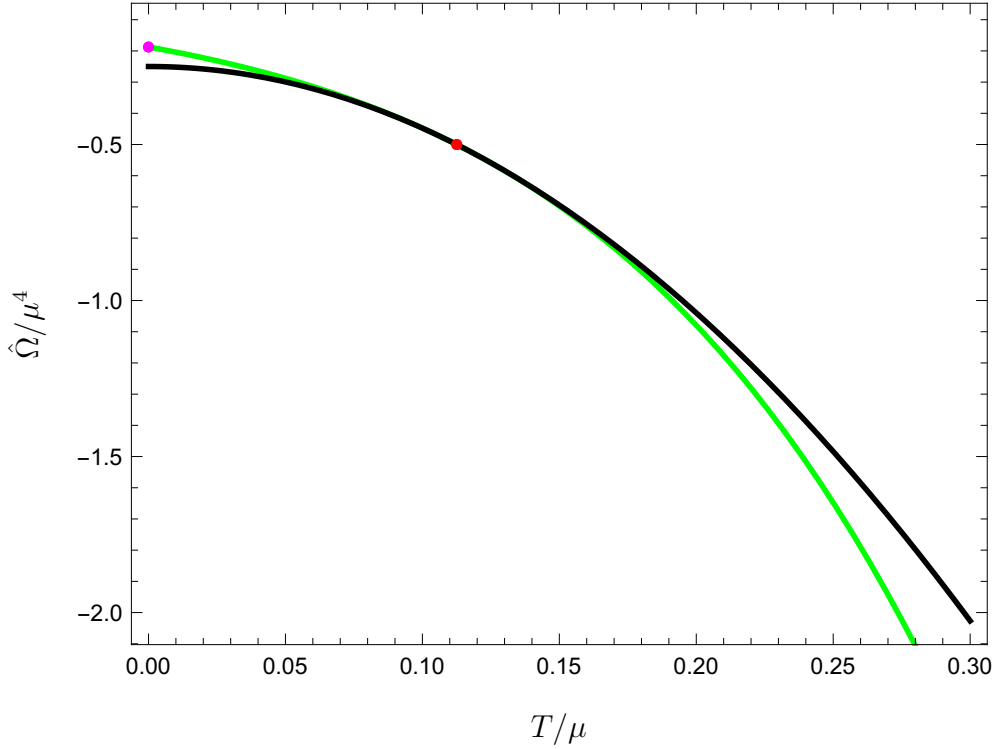}
\end{center}
  \caption{The Gibbs free energy density of the disordered phase (the green curve)
  and the ordered phase (the black curve) of the charged $\caln=4$ SYM plasma. The red dot
  indicates the critical temperature \eqref{tcrit}; the magenta dot represents the extremal RN black hole.
} \label{fa}
\end{figure}

There are two different phases of the STU black holes \eqref{thermo} when all the $U(1)$
chemical potentials $\mu_a$ are equal:
\begin{itemize}
\item The obvious RN-phase where we take
\begin{equation}
\kappa_1=\kappa_2=\kappa_3\,,
\eqlabel{rnvev}
\end{equation}
resulting in the equation of state \eqref{omsysi}. In this phase all the
gravitational bulk scalars are trivial, \ie $X_a\equiv 1$.
\item The ordered phase of \cite{Buchel:2025cve}:
\begin{equation}
\kappa_1=\kappa_2=\frac{1}{\kappa_3}\equiv \kappa\,.
\eqlabel{ovev}
\end{equation}
From \eqref{ovev}, we identify
\begin{equation}
\kappa=\frac {\mu^2}{8\pi^2 T^2}\,,\qquad T_0^2=\frac{4\mu^2 T^2}{8\pi^2 T^2+\mu^2}\,,
\eqlabel{kt0}
\end{equation}
resulting in, see \eqref{thermo},
\begin{equation}
\begin{split}
&\Omega_{ordered}=-\frac{N^2}{32\pi^2}\ \mu^2(\mu^2+8\pi^2 T^2)\,,\qquad \cale_{ordered}=-3\Omega_{ordered}\\
&s_{ordered}=\frac{N^2}{2}\ \mu^2 T\,,\qquad \rho_1^{ordered}
=\rho_2^{ordered}=\frac{N^2}{16\pi^2}\ \mu^3\,,\qquad \rho_3^{ordered}=\frac{N^2}{2}\
\mu T^2\,.
\end{split}
\eqlabel{sf}
\end{equation}
Additionally, since in the ordered phase the gravitational bulk scalars are nontrivial,
there is an expectation value of the corresponding dimension-2 operator of the SYM:
\begin{equation}
\frac{\calo_2^{ordered}}{\pi^2 T_0^2}\equiv \lim_{u\to 0} \frac{dX_1}{du} = \frac{64\pi^4 T^4-\mu^4}{24\pi^2\mu^2 T^2}\,.
\eqlabel{ovevsss}
\end{equation}
\end{itemize}

In fig.\ref{fa} we compare the Gibbs free energy density $\hat\Omega\equiv \frac{8\pi^2}{N^2}\Omega$ of
the ordered phase (the black curve) and the
disordered phase  --- with the RN black hole gravitational dual \eqref{omsysi} --- (the green curve)  .
The magenta dot represents the extremal $T=0$ limit, and the red dot represents the onset of the
hydrodynamic instability identified in \cite{Gladden:2024ssb}.
In the vicinity of the critical temperature, see \eqref{tcrit},
\begin{equation}
\frac{8\pi^2}{N^2}\ \frac{\Omega_{ordered}-\Omega_{RN}}{\mu^4}\ =\ \frac{32\pi^3\sqrt{2}}{27}\left(
\frac{T-T_{crit}}{\mu}\right)^3
+\calo\left((T-T_{crit})^4\right)\,,
\eqlabel{diffom}
\end{equation}
implying \eqref{omcomp}.

Once we have the analytic expression for the equation of state
of the ordered phase \eqref{ovev}, we can verify whether this phase is
thermodynamically stable\footnote{I would like
to thank A.Starinets and P.Kovtun for pointing this out.}. 
One of the stability conditions of STU states \eqref{thermo} is
\cite{Gladden:2024ssb}
\begin{equation}
2-\sum_a \kappa_a-\prod_b \kappa_b >0\,,
\eqlabel{stab}
\end{equation}
which is always violated in the ordered phase \eqref{ovev}.
Whether or not the hydrodynamic instability associated
with this thermodynamic instability survives at nonlinear level
is an open question. It is an open question as to what is the end point of
this potential instability.

\section*{Acknowledgments}
I would like to thank Pavel Kovtun and Andrei Starinets for valuable discussions.
Research at Perimeter Institute is supported by the Government of Canada through Industry
Canada and by the Province of Ontario through the Ministry of
Research \& Innovation. This work was further supported by
NSERC through the Discovery Grants program.

\bibliographystyle{JHEP}
\bibliography{n4rn2}

\end{document}